\documentstyle[12pt]{article}
\begin{document}

\title{\bf Exact trace formulae for a class of
              one-dimensional ray-splitting systems}
\author
{Y. Dabaghian, R. V. Jensen and R. Bl\"umel \cr
Department of Physics, Wesleyan University, \cr
Middletown, CT 06459-0155
}
\date{\today}
\maketitle
\begin{abstract}
Based on quantum graph theory we establish that
the ray-splitting trace formula proposed by
Couchman {\it et al.}
(Phys. Rev. A {\bf 46}, 6193 (1992)) is exact
for a class of one-dimensional ray-splitting systems.
Important applications in
combinatorics are suggested.
\end{abstract}

PACS: 05.45.+b,03.65.Sq,72.15.Rn

\section{Introduction}
Gutzwiller's trace formula, established in the late 1960s and early 1970s
states that it is possible to obtain the level density of a bounded
Hamiltonian dynamical system with semiclassical accuracy, based entirely on
the information provided by its classical periodic orbits \cite{Gutz,Gbook}.
According to Gutzwiller the density of energy states can be written as a sum
over prime periodic orbits and their repetitions,
\begin{equation}
\rho \left( E\right) =
\sum_n\, \delta(E-E_n)=\bar\rho(E) + 
\frac{1}{ \pi \hbar} {\rm Re}\,
\sum_{p}T_{p}(E)\sum_{\nu=1}^{\infty }A_{p\nu }(E)
e^{i\nu \left[ \frac{S_{p}\left( E\right)}{\hbar}+\varphi_{p}(E)\right] },
\label{a2}
\end{equation}
where $\bar\rho(E)$ is the average density of states,
$S_{p}$ is the classical action of the prime periodic orbit $p$,
\begin{equation}
T_{p}=\frac{\partial S_{p}(E)}{\partial E}
\label{period}
\end{equation}
is its period and $\varphi _p(E)$ is its Maslov phase.
Gutzwiller derived the pre-exponential factors $A_{p\nu}$ in semiclassical
approximation, expressing them in terms of the stability properties of the
corresponding periodic orbits.

The orbits used to construct the sum (\ref{a2}) are obtained at a given value
of the energy $E$. On the other hand it is known that as the energy of a
generic, nonhyperbolic system changes, the structure of the phase space
changes and with it the set of periodic orbits.
This phenomenon is called ``phase-space metamorphosis'' \cite{meta}.
Phase-space metamorphosis, in general, is accompanied by the creation and
destruction of periodic orbits giving rise to the interesting phenomenon of
ghost orbits \cite{Ghost}. Therefore, in general, the sum in (\ref{a2}) will
change as a function of $E$ in the sense that it may acquire or lose certain
terms. However, apart from hyperbolic systems\cite{GH}, there exists an
interesting class of systems, which are free of such metamorphoses of the
phase space. These are the scaling systems, for which the action functional
$S_p(E)$ for any periodic orbit $p$ decouples into the product of an
energy-dependent part $f(E)$ and the ``reduced action'' $S_{p}^{0}$, which
depends only on the geometry of the periodic orbit,
\begin{equation}
S_p(E) =f(E) S_p^{0}.
\label{a3}
\end{equation}
Examples of such systems are various billiard systems, and also 
(with suitable definitions of scaling parameters)
the hydrogen atom in strong electric and/or
magnetic fields \cite{Gbook,FW,FF}. For such 
systems neither the geometry of the phase
space nor the geometry of the set of periodic orbits change with
energy. Therefore the structure of the sum (\ref{a2}) can be defined
once and is valid for all values of $E$. In such cases it is interesting to
investigate the relationship between the fixed set of periodic orbits (the
periodic orbit spectrum) and the quantum energy spectrum.

Direct derivation of Gutzwiller's trace formula, as presented
originally by Gutzwiller, is based on
the saddle point approximation.
This implies that this formula is meant
to work only semiclassically, i.e.
to predict only the highly
excited energy levels with semiclassical accuracy.
Indeed, for generic billiard domains,
Gutzwiller's formula is not exact \cite{Gutz}.
However, in certain special cases
(\ref{a2}) is known to predict the entire
energy spectrum exactly. An example is the
harmonic oscillator.
Moreover, exact ``Gutzwiller-like'' trace formulae
do exist.
A theorem by Anderson and Melrose \cite{AM} states that for any billiard
there exists a set of pre-exponential factors $A_{\alpha}$, which makes the
relationship (\ref{a2}) exact. Other exact Gutzwiller-like trace formulae are
obtained in the context of quantum graph theory\cite{QGT1,QGT2,QGT3}.
We show that ray-splitting systems \cite{Couch}
provide further examples of
exact Gutzwiller-like trace formulae.
A specific example,
a scaling one-dimensional step billiard,
closely related to quantum graphs, is
presented in the following section together with its
generalized ray-splitting Gutzwiller formula.
In Section III we
present a proof for the exactness of the ray-splitting
Gutzwiller formula. In Section IV we define the
ray-splitting zeta function and relate it to
cycle expansion techniques.
In Section V we make use of the exactness
of the ray-splitting Gutzwiller
formula to prove a nontrivial combinatorial
identity. In Section VI we discuss our results and
conclude the paper.

\section{The model}
In this section we study the spectrum and the generalized
ray-splitting Gutzwiller formula for
the one-dimensional scaling
step billiard\cite{Couch,RS1,RS2,RS3,RS4} (Fig.~1)
\begin{equation}
V(x) =\cases{
0, &for $0<x\leq b$, \cr
V_0=\lambda E, &for $b<x<1$,\cr}
\label{a4}
\end{equation}
where $\lambda$ is the scaling constant and $E$ is the
energy of the system. In this paper we focus on the case
$\lambda<1$. First results on this model were presented
in \cite{RS4} in the context of generalized Poisson
formulae in ray-splitting systems.
Despite its formal simplicity (\ref{a4})
can be used to illustrate many physical and
mathematical methods and ideas connected with
the ray-splitting approach\cite{Couch,RS1,RS2,RS3,RS4}.
We work
in units such that $\hbar=1$, the width of the potential well
is 1 and the mass of the
quantum particle is 1/2. Defining $k=\sqrt{E}$ and
$\kappa =\beta k$, where
\begin{equation}
\beta =\sqrt{1-\lambda },
\end{equation}
it is elementary to obtain the exact quantum mechanical equation
\begin{equation}
\cos(kb)\sin[\kappa(1-b)] +{\kappa\over k}\sin(kb)\cos[
\kappa(1-b)] =0
\label{z1}
\end{equation}
for
the energy levels $E_n$ of the system. They
are determined by the
roots $k_n$ of (\ref{z1}) according to $E_n=k_n^2$.

It is more convenient to write (\ref{z1}) in the form
\begin{equation}
\sin(k\omega_1)-r\sin(k\omega_2)=0
\label{a6}
\end{equation}
with
\begin{equation}
\omega_1=l_1+l_2,\ \ \ \omega_2=l_1-l_2,\ \ \ l_1=b,\ \ \
l_2=\beta (1-b)
\label{ll}
\end{equation}
and the reflection coefficient
\begin{equation}
r={1-\beta\over1+\beta}.
\label{reflco}
\end{equation}
In general the two frequencies
$\omega_1$ and $\omega_2$ in (\ref{a6}) are not
rationally related. Therefore
(\ref{a6}) connects the physical problem
of a scaling step potential with the mathematical theory of
almost periodic functions\cite{HBohr}. This means that all our exact
results on the spectrum of the scaling step potential (\ref{a4})
can be interpreted as theorems on the roots of doubly periodic
functions.

Solving (\ref{a6}) numerically,
it is easy to obtain a large number of roots
for studying statistical properties
of the quantum energy levels as well as
the relationships to the classical
periodic orbits of the system. The latter
goal is achieved by computing the
Fourier image of the density of states defined as
\begin{equation}
F(s) \, =\, \sum_{j=1}^{\infty}\, e^{-isk_{j}}.
\label{a7}
\end{equation}
According to the Gutzwiller trace formula (\ref{a2}), the Fourier transform
provides a convenient tool for studying the orbit spectra of dynamical
systems, since it produces pronounced peaks at those values of $s$ that
correspond to the actions of classical periodic orbits. In the case of the
scaling system (\ref{a4}), the actions $S_{p}$ in (\ref{a2}) are proportional
to $k$,
\begin{equation}
S_{p}\left( E\right) =\int_{p}k\left( x\right) dx=S_p^0\cdot k,
\label{a8}
\end{equation}
and hence one expects the Fourier transform (\ref{a7}) of (\ref{a2}) to
produce a $\delta$-peak at $s_{p\nu}=\nu S_p^0$ for every primitive periodic
orbit $p$ and its repetitions $\nu$.

The result of the numerical evaluation of the sum (\ref{a7}) for this system
is presented in Fig.~2. It shows a large number of narrow peaks. Most of them
do not correspond to the standard (Newtonian) periodic orbits. This is
immediately clear since in the case of the potential (\ref{a4}) there exists
only a single primitive Newtonian periodic orbit at any value
of the energy above the potential step (see Fig.~1).

The extra peaks in Fig.~2 are due to non-Newtonian periodic
orbits\cite{Couch,RS1,RS2,RS3,RS4}. They correspond to the non-Newtonian
reflections off the sharp ray-splitting step. Together with the Newtonian
orbits the non-Newtonian orbits account for every single peak in Fig.~2 for
arbitrary values of the parameters $\lambda$ and $b$. Numerical computations
indicate that the maxima
of $F(s)$ converge to $\delta$-peaks in the limit when
the number of roots included in the sum (\ref{a7}) tends to infinity. This,
in turn, suggests that there exists an exact formula of the type
(\ref{a2}). That this is indeed the case is proved in Sect. III below.

A natural generalization of Gutzwiller's trace formula, which includes the
contributions from the non-Newtonian ray-splitting orbits, was obtained
previously in \cite{Couch}. Speaking in terms of the step-potential
(\ref{a4}), instead of just a single orbit bouncing
between $x=0$ and $x=1$, a generic
orbit may now be reflected off or
transmitted through the ray-splitting boundary at
$x=b$ any number of times in
arbitrary sequence. As a result, the set of
{\it primitive non-Newtonian orbits}
becomes infinite. In a one-dimensional system
the numbers of these reflections and transmissions
are the only characteristics
of the orbits, and therefore any orbit can be
characterized uniquely and completely by a binary
sequence of symbols $\cal{L}$ and $\cal{R}$ that keep track
of each reflection off the left ($\cal{L}$) or the
right ($\cal{R}$) wall of the potential well.
The corresponding generalized Gutzwiller
sum includes all
primitive Newtonian and non-Newtonian
periodic orbits and
their repetitions,
\begin{equation}
\rho (E) =\bar\rho(E) + 
\frac{1}{ \pi} {\rm Re} \sum_{p}T_{p}\sum_{\nu
=1}^{\infty }\left[(-1)^{\chi(p)}t^{2\tau(p)}
r^{\sigma(p)}\right]^{ \nu }
\, e^{i\nu S_{p}},
\label{a9}
\end{equation}
where $\bar\rho = (b+\beta(1-b))/(2\pi k)$ is the average level density,
$r$, defined by (\ref{reflco}), is the quantum reflection
coefficient, $t=\sqrt{1-r^2}$ is the transmission coefficient,
$\sigma(p)$ and $2\tau(p)$ are the number of
reflections and transmissions of
the primitive orbit $p$ at the potential step,
and $\chi(p)$ counts the total number of times
the orbit reflects off the walls and off the
potential step to the right of the ray-splitting
boundary. Note
that $(-1)^{\chi(p)}=e^{i \varphi_{p}}$ and explicitly
defines the Maslov phase in (\ref{a2}).

If we denote the actions of the
shortest orbits ($\cal{R}$ and $\cal{L}$) by
$S_{\cal{R}}$ and $S_{\cal{L}}$
respectively, the action $S_{p}$ of any orbit
can be expressed as a sum
\begin{equation}
S_{p}=n_{\cal{L}}S_{\cal{L}}+n_{\cal{R}}S_{\cal{R}},
\label{action}
\end{equation}
for certain integers
$n_{\cal{L}}$ and $n_{\cal{R}}$
(generally different from $\sigma$ and
$\tau$).
The level density (\ref{a9})
contains only even powers of the transmission
coefficient $t$, because every
periodic orbit transmits an even number of
times through the ray-splitting boundary.

\section{Exactness of the ray-splitting trace formula}
Using quantum graph
theory \cite{QGT1,QGT2,QGT3}, it is possible to show that the
expression (\ref{a9})
is exact. We prove this below after presenting some
basic ideas of quantum graph theory.

>From the perspective of quantum graph theory,
the quantization of a particle in
the potential (\ref{a4}) is treated
as a scattering problem on the graph
\begin{equation}
{\bullet}_{1} {\line(1,0){50}}{\bullet}_2 {\line(1,0){50}}
{\bullet}_3
\label{a10}
\end{equation}
with three vertices and
two bonds described by the connectivity matrix
\cite{QGT1,QGT2,QGT3}
\begin{equation}
C= \pmatrix{0&1&0 \cr 1&0&1 \cr 0&1&0}.
\end{equation}
On every bond connecting
vertices $i$ and $j$, one defines a free particle wave
function $\psi _{ij}$, which
satisfies the following vertex conditions:
\begin{eqnarray}
\psi _{ij}(\xi=0)=\varphi _{i},\ \ \
\psi _{ij}(\xi=L_{ij})=\varphi_{j},
\end{eqnarray}
where $\xi$ is the coordinate
along a particular bond of length $L_{ij}$, so
that the wave functions on
different bonds match on every vertex. The general
solution satisfying the vertex conditions is
\begin{equation}
\psi _{ij}(\xi) =\frac{\varphi _{i}C_{ij}}{\sin[k\beta
_{ij}L_{ij}]}\sin[ k\beta _{ij}\left( L_{ij}-\xi\right)] +\frac{\varphi
_{j}C_{ij}}{\sin[ k\beta_{ij}L_{ij}]}\sin[ k \beta_{ij}\xi].
\end{equation}
The presence of the
coefficients $\beta_{ij}$ allows us to generalize the
formalism developed
in \cite{QGT1,QGT2,QGT3}. For the derivatives we have the
continuity conditions
\begin{eqnarray}
\sum_{j<i}C_{ij}\psi_{ji}^{\prime }(\xi =L_{ij})=
\sum_{j>i}C_{ij}\psi_{ij}^{\prime }(\xi =0).
\end{eqnarray}
For the case of the potential (\ref{a4}),
\begin{equation}
\psi_{12}(\xi) =\frac{\varphi _{2}}{\sin( kb)}\sin(k\xi),
\end{equation}
\begin{equation}
\psi_{23}(\xi) =
\frac{\varphi_{2}}{\sin[k\beta \left( 1-b\right)]
}\sin[ k\beta \left( 1-b-\xi\right)] .
\end{equation}
The matching and continuity conditions at vertex 2 result in
\begin{equation}
\beta \tan(kb)+\tan[k\beta(1-b)]=0
\end{equation}
or
\begin{equation}
\sin[k(l_{1}+l_{2})]-r\sin[k(l_{1}-l_{2})]=0,
\label{transc}
\end{equation}
which is the same as (\ref{a6}). The lengths
$l_1$ and $l_2$, defined in (\ref{ll})
turn out to be the weighted bond lengths.

The same quantization
condition can be obtained from considering the
scattering process at
every vertex of the graph (\ref{a10}). The vertex
scattering matrix is given by
\begin{equation}
\sigma_{jiij'}^{(i)}=g_{jj'}^{(i)}
C_{ji}C_{ij^{\prime }},
\label{a11}
\end{equation}
where the index $i$ refers to the vertex under consideration,
$g_{jj'}^{(i)}$ are coefficients that depend on the physics
of the scattering at the vertex $i$ and $C_{ij}$ are the
matrix elements of the connectivity matrix determining
the geometry of the graph.
At the ``dead end'' vertices 1 and
3 of the graph (\ref{a10}) we have $\sigma^{(1)}_{j11j'}=
\sigma^{(3)}_{j33j'}=-1$. For the central vertex it is easy to show that
\begin{equation}
\sigma^{(2)}_{j22j'}=
\pmatrix{r & t \cr t & -r}.
\label{sigma}
\end{equation}
The graph scattering matrix, describing the graph as a whole, is given by
\begin{equation}
S= \pmatrix {0 & -D \cr D\sigma ^{(2)} & 0},
\label{a12}
\end{equation}
where the matrix
\begin{equation}
D= \pmatrix{e^{il_{1}k}&0 \cr 0 & e^{il_{2}k}}
\end{equation}
accounts for the phases accumulated along the bonds.
The quantization condition\cite{QGT1,QGT2,QGT3},
\begin{equation}
\det \left( 1-S\right) =0,
\label{quaco}
\end{equation}
results in
\begin{equation}
e^{2i\left(l_{1}k+l_{2}k\right)}-r\left(e^{2il_{1}k}-e^{2il_{2}k}\right)=1.
\end{equation}
This is the same as (\ref{transc}).

With the help of
\begin{equation}
\ln \det(1-S)=-\sum_{n=1}^{\infty} {1 \over n} Tr \left(S^{n}\right),
\end{equation}
the quantization condition (\ref{quaco}) can be written alternatively as a
sum over the periodic orbits of the graph. Indeed, since the scattering
matrix is defined geometrically using the graph connectivity matrix, its
indices correspond to the vertices $i$ and $j$ connected by a bond if the
matrix element $C_{ij} \neq 0$. The trace of the $n$-th power of this
matrix is defined on the set of all the possible cyclical $n$-bond
sequences. Using (\ref{a11}) and (\ref{a12}) we obtain 
$Tr (S^{2n+1})=0$ and 
\begin{equation}
Tr \left(S^{2n}\right)=2\sum_{n_{\cal{L}}+n_{\cal{R}}=n}
(-1)^{\chi}r^{\sigma}t^{2\tau}e^{2ikL_{n}},
\end{equation}
where $n_{\cal{L}}$ and $n_{\cal{R}}$ give the number of times the left
($\cal{L}$) and the right ($\cal{R}$) bonds of (\ref{a10}) occur in the
sequence, $L_{n}=n_{\cal{L}}l_{1}+n_{\cal{R}}l_{2}$, $\sigma$ is the number
of reflections from the middle vertex and
$2\tau$ is the number of transmissions through it. Since the reflection
coefficient coming from the scattering matrix (\ref{sigma}) can be positive
or negative, the factor $(-1)^\chi$ is needed to keep track of how many times
it appears with a minus sign. It also keeps track of how many times
a given orbit scatters off the walls.

There are two possibilities for $L_n$. Either it corresponds to
a prime periodic orbit,
or it corresponds to a bond sequence that
retraces itself $\nu$ times. In this case
$L_{n}$ is $\nu$
times the length of a single (primitive) traversal $L_{p}$, and the
pre-exponential factor is the $\nu$-th power of the factor corresponding to
the shortest closed bond sequence. Alternatively, these closed bond sequences
can be viewed as periodic orbits traced by a particle moving on the
graph.

On the other hand, from the analytical properties of
$\det[1-S(E)]$ \cite{QGT3},
it is easy to relate it to the spectral counting function $N(E)$,
\begin{equation}
N(E)=\bar N_{\rm W}(E)-{1\over 2}+\frac{1}{\pi} {\rm Im} 
\sum_{n=1}^{\infty}\frac{1}{n}Tr
\left(S^{n}\right),
\end{equation}
where $\bar N_{\rm W}(E)$ is the average spectral 
staircase function (the Weyl
term).
Putting all these ingredients together, one arrives at the trace
formula (\ref{a9}).

The exactness of (\ref{a9}) can be understood on the basis of the dramatic
increase of the number of the primitive non-Newtonian orbits included in
(\ref{a9}).
As the orbit length $L_{p}$ in (\ref{a9}) increases,
non-Newtonian orbits proliferate exponentially, providing additional
information about the structure of the potential $V(x)$.

It is interesting to compare (\ref{a9}) with the results produced by
Gutzwiller's trace formula without ray splitting. For the step potential
(\ref{a4}) there exits only one Newtonian periodic orbit with
classical action $S_N$ which bounces
between the left and the right walls of the well (Fig.~1).
Therefore (\ref{a9}) predicts
\begin{equation}
\rho\left(E \right)=\bar\rho(E)+\frac{T_{N}}{\pi}\sum_{\nu =
1}^{\infty}\cos(\nu S_N)=
T_{N}\, \sum_{m=-\infty}^{\infty}\delta \left( S_N-2\pi m\right),
\label{a15a}
\end{equation}
where 
\begin{eqnarray}
S_N=2kb+2k\beta (1-b)=S^{0}_{N}k,
\label{newtact}
\end{eqnarray}
$T_{N}=\partial S_N/\partial E$ and  
$\bar\rho=T_N/2\pi$.
Since only a single periodic orbit contributes,
(\ref{a15a}) predicts
a periodic spectrum for a particle in the step potential
(\ref{a4}). Moreover,
the Fourier image
\begin{equation}
F\left( s\right) =\int \rho(E)\, e^{-isk}\, dE = 
\left[S_N^0\right]^2\, 
\sum_{m=-\infty}^{\infty }\delta \left[m
S^{0}_{N}-s\right]
\end{equation}
of the 
density of states produces $\delta$-peaks
at integer multiples of the reduced 
action $S^{0}_{N}$ of the (only) Newtonian
orbit of the system. Figure 3 shows that the exact spectrum of the problem is
not periodic, which illustrates that the trace 
formula (\ref{a2}) ((\ref{a9}), respectively) without
non-Newtonian orbits predicts a wrong energy level distribution.

A straightforward generalization of the ideas and procedures discussed above
provides a proof of the exactness of (\ref{a9}) for the whole class of
$N$-step scaled potentials
\begin{equation}
V(x)=V_{i}=\lambda_{i}E,\ \ \ \ \ b_{i-1}<x<b_{i},\ \ \ i=1,\ldots,N,
\label{genstep}
\end{equation}
where $b_0=0$, $b_N=1$ and
$\lambda_i$ is the scaling coefficient for the $i$-th
interval $[b_{i-1},b_i]$.

\section{Ray-splitting zeta function}
Despite its exponentially decreasing terms, (\ref{a9}) converges only
conditionally due to the exponential
proliferation of the non-Newtonian orbits. Therefore, one should specify a
physically meaningful way of partial summation for this series. In practice,
one could certainly consider the shortest periodic orbits in order to get an
approximation for (\ref{a9}). Figure 3 presents the contribution of the
43 shortest periodic orbits. This corresponds to including all periodic orbits
up to binary code length 7 (e.g. $\cal{LRRLLLL}$). Figure 3 shows that the
peaks give a very accurate representation of the actual positions of the
roots.

Since (\ref{a9}) is a geometric series with respect to the repetition
index $\nu$, this part of the
summation can be performed immediately, yielding
\begin{equation}
\rho(E) =\bar\rho(E)+
{1\over \pi}\, {\rm Re} \sum_{p}\, T_{p}
\frac{(-1)^{\chi(p)}t^{2\tau(p)}r^{\sigma(p)}e^{iS_{p} \left(E\right)}}
{1-(-1)^{\chi(p)}t^{2\tau(p)}r^{\sigma(p)}e^{iS_{p}\left( E\right)}}.
\label{a16}
\end{equation}
Using the relation (\ref{period}), the density of states can be written as
\begin{equation}
\rho(E)=\bar\rho(E) - {1\over\pi}{\rm Im}
\frac{\partial }{\partial E}\ln Z\left(E\right),
\end{equation}
where
\begin{equation}
Z(E)=
\prod_{p}\left[ 1-(-1)^{\chi(p)}t^{2\tau(p)}r^{\sigma(p)}
e^{iS_{p}(k)}\right]
\label{a17}
\end{equation}
is an analog of the Fredholm determinant associated with the ray-splitting
system (\ref{a4}) and the sum (\ref{a9}), considered as a function of the
coefficients $r$ and $t$. One can consider a cycle expansion \cite{b10}
of the product (\ref{a17}) in powers of $r$ or $t$. A natural choice for the
expansion variable would
be the smaller one of $r,t$.
Physically, this asymmetry determines whether reflection or transmission
is the dominant process.

\section{Combinatorics}
In quantum graph theory the representation of the quantum
level density in the form of a Gutzwiller-like trace formula
is exact\cite{QGT1,QGT2,QGT3}. In special cases both the level
density and the trace formula can be evaluated analytically
and give rise to
combinatorial identities.
This idea was
successfully implemented by Schanz and Smilansky
who
obtained a host of new and nontrivial combinatorial identities
\cite{combi}.
Additional identities are generated whenever
(\ref{transc}) can be solved analytically.
We illustrate the method by choosing $\beta$ such that $l_{1}=b$ and
$l_{2}=\beta(1-b)=b$ in (\ref{transc}).
In this case (\ref{transc}) becomes
$\sin(2kb)=0$, solved by $k_n=n\pi/(2b)$. The corresponding periodic
level density can be obtained directly using the conventional Poisson
formula,
\begin{equation}
\rho(E)=\sum_{n=1}^{\infty} \delta \left(E-\frac{\pi^2
n^{2}}{4b^{2}}\right)=
\frac{b}{\pi k}\, \sum_{m=-\infty}^{\infty}\, e^{i m 4bk}.
\label{rex}
\end{equation}
It is interesting that the arguments of the exponents in (\ref{rex}) coincide
with the actions of the repetitions of the Newtonian orbit, $S_N=4bk$. This
coincidence is due to the special choice of parameters, $l_{1}=l_{2}=b$
assumed in this case.

Alternatively, the level density (\ref{rex}) can be expressed via
(\ref{a9}). Equating the pre-factors of terms with the same action
results in the following sum rule
\begin{equation}
1 = {1\over T_N} \, \sum_{p\nu}T_{p}
\left[(-1)^{\chi(p)}t^{2\tau(p)}r^{\sigma(p)}\right] ^{\nu }.
\label{sr}
\end{equation}
Here $T_N=2b/k$ and the sum on the right-hand side is over all
periodic orbits, Newtonian and non-Newtonian, that add up
to the same multiple of the Newtonian action $S_N$.

The sum rule (\ref{sr}) can be recast into a combinatorial
theorem on the set $W_{\Lambda}$ of cyclically non-equivalent
binary codes
of even length $\Lambda=2M$
over the symbols $\cal{L}$ and $\cal{R}$ in the following way.
(i) For every word $w\in W_{\Lambda}$ compute
the primitive time $T_w$ defined as $T_w=M/\nu_w$, where
$\nu_w$ is the number of repetitions of the shortest
sub-code in $w$.
(ii) Scan each word $w\in W_{\Lambda}$ and assign
\begin{equation}
w\rightarrow (-1)^{\alpha_w}\, (r^2)^{\beta_w}\, (t^2)^{\gamma_w}
\label{abc}
\end{equation}
according to the substitutions
\begin{equation}
{\cal LR}\rightarrow t,\ \
{\cal RL}\rightarrow t,\ \
{\cal LL}\rightarrow r,\ \
{\cal RR}\rightarrow -r.
\end{equation}
Then, with $r^2+t^2=1$, we have
\begin{equation}
\sum_{w\in W_{\Lambda}}\, T_w\,
(-1)^{\alpha_w}\, (r^2)^{\beta_w}\, (t^2)^{\gamma_w}
\ =\ 1,
\label{cth}
\end{equation}
which is equivalent to the sum rule (\ref{sr}).
Stated differently, (\ref{cth}) is the same as
\begin{equation}
\sum_{w\in W_{\Lambda},\beta_w=\beta}\, (-1)^{\alpha_w}\, T_w =
\left(\matrix{M\cr\beta\cr}\right).
\end{equation}

\section{Discussion and conclusions}
Since Atle Selberg discovered his famous trace formula in 1947 \cite{Sel},
more exact trace formulae were found. There are many cases in which the
geometrical information contained in the set of closed geodesics (periodic
orbits) can be used to reconstruct the spectrum exactly. Gutzwiller provided
a physical theory which parallels these results. His formula points the way
to establish {\it approximate} relationships, which involve physical rather
than geometrical concepts.
In particular, Gutzwiller uses the semiclassical saddle-point approximation,
valid under certain physical conditions, in order to derive the
pre-exponential factors in (\ref{a2}) in a form, which is valid for a wide
class of dynamical systems.

However, Gutzwiller's theory does not imply that these sums are necessarily
approximate. There exist different approaches to establish exact
relationships between the spectra of operators and the spectra of periodic
orbits. The above method, based on the intuitively clear idea of ray
splitting, provides an example which is physical, lies outside of the scope
of Gutzwiller's approach, and is exact.

This result immediately suggests many important applications. First, from the
mathematical point of view, one can derive statements about the behavior of
the zeros of a wide class of almost periodic functions. Second, the exactness
of (\ref{a9}) provides a convenient way to prove many combinatorial
identities, expressible in terms of periodic orbits in (\ref{a4}). Lastly, it
provides a non-trivial way to obtain Feynman's path integrals in a
well-defined limit. It is a natural idea to approximate arbitrary
one-dimensional potentials by a step-like profile such as (\ref{genstep}),
for which (\ref{a9}) is exact. Taking the limit in which the size of the
steps tends to zero, one can approximate the shape of any smooth potential with
any accuracy. As the number of steps increases, the sum over the
non-Newtonian orbits leads to Feynman's path integral.

Y.D. and R.B. gratefully acknowledge financial support by
NSF grants PHY-9900730 and PHY-9984075; Y.D. and
R.J by NSF grant PHY-9900746.



\centerline{\bf Figure Captions}
\bigskip\noindent
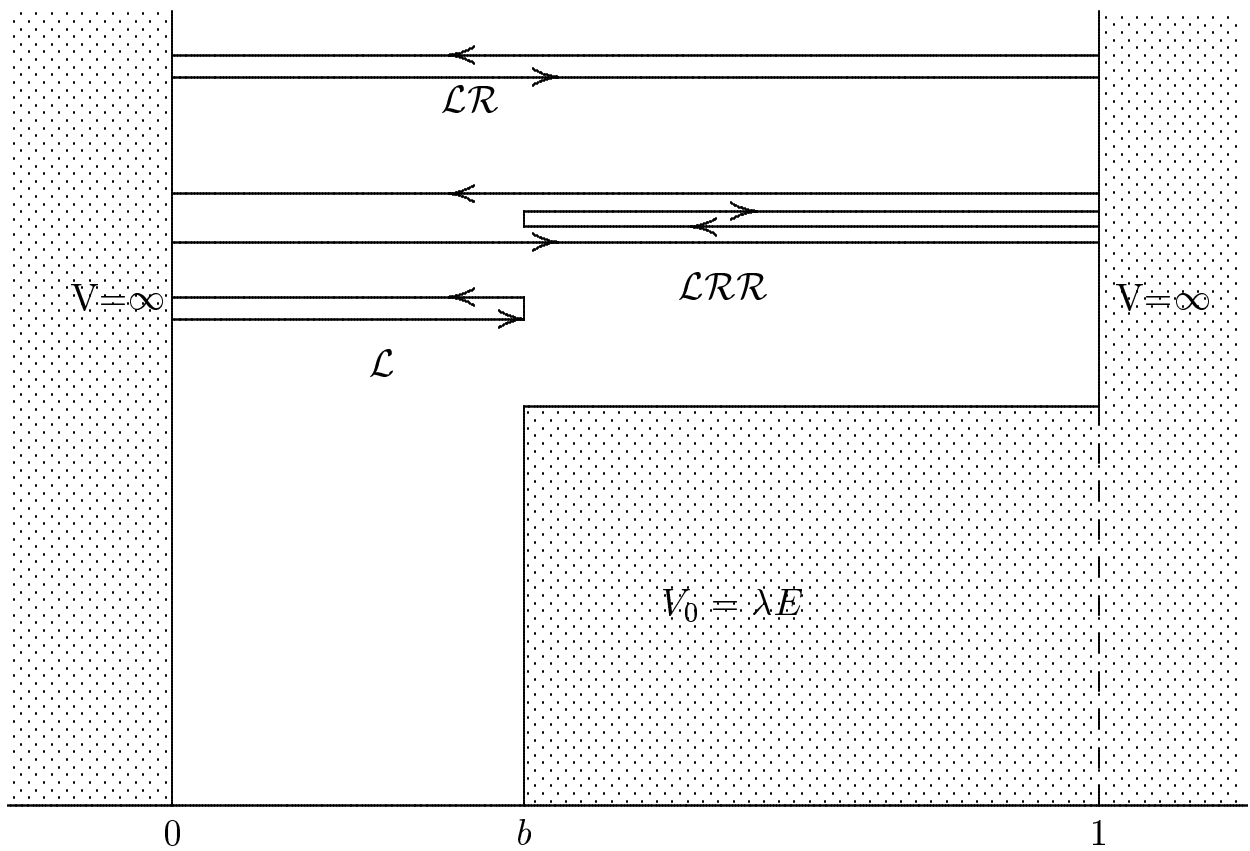
{\bf Fig.~1:} Step potential with non-Newtonian orbits
$\cal{L}$, $\cal{LRR}$ and Newtonian orbit $\cal{LR}$.

\bigskip\noindent
{\bf Fig.~2:} Fourier transform (\ref{a7}) of the density of states of the
step potential (\ref{a4}) with $b=0.7$ and $\lambda=1/2$. About 10,000 states
are included in the sum (\ref{a7}). Sharp peaks in the transform are located
precisely at the actions of the Newtonian and
non-Newtonian orbits.

\bigskip\noindent
{\bf Fig.~3:} Contribution of the 43 shortest Newtonian and non-Newtonian
periodic orbits (up to binary code length 7) to the density of states
of the step potential shown in Fig.~1. The exact energy eigenvalues ($+$)
are close to the locations of the peaks. For comparison the energy levels
predicted by the Newtonian orbits alone are also shown ($\times$).


\begin{thebibliography}{99}

\bibitem{Gutz}
             M. C. Gutzwiller,
             J. Math.\ Phys. {\bf 8}, 1979 (1967);
             {\bf 10}, 1004 (1969); {\bf 11}, 1791 (1970);
             {\bf 12}, 343 (1971).

\bibitem{Gbook}
             M. C. Gutzwiller, {\it Chaos in Classical and Quantum
             Mechanics} (Springer, New York, 1990).

\bibitem{meta}  Y. C. Lai, C. Grebogi,
              R. Bl\"{u}mel, and M. Ding, Phys. Rev. A {\bf 45},
              8284 (1992).

\smallskip\noindent
\bibitem{Ghost} M. Ku\'s, F. Haake, and D. Delande,
     Phys.\ Rev.\ Lett.\ {\bf 71},
 2167 (1993).

\bibitem{GH}  J. Guckenheimer and P. Holmes, {\it Non-Linear
              Oscillations, Dynamical Systems, and
              Bifurcations of Vector Fields} (Springer,
              New York, 1983).

\bibitem{FW}  H. Friedrich and D. Wintgen, Phys. Rep. {\bf 183},
              37 (1989).

\bibitem{FF} H. Friedrich, {\it Theoretical Atomic Physics}
  (Springer, Berlin 1998).


\bibitem{AM}  K. G. Anderson and R. B. Melrose,
              Inv. Math. {\bf 41},
              197 (1977).

\bibitem{QGT1} T. Kottos and U. Smilansky, Phys. Rev. Lett.
              {\bf 79}, 4794 (1997).

\bibitem{QGT2} H. Schanz and U. Smilansky, Phys. Rev. Lett. {\bf 84},
              1427 (2000).

\bibitem{QGT3} T. Kottos and U. Smilansky, Annals of Physics
              {\bf 274}, 76 (1999).

\bibitem{Couch} L. Couchman, E. Ott, and T.M. Antonsen, Jr.,
Phys. Rev. A {\bf 46}, 6193 (1992).

\bibitem{RS1} R. E. Prange, E. Ott, T. M. Antonsen, B. Georgeot,
              and R. Bl\"umel, Phys. Rev. E {\bf 53}, 207 (1996).

\bibitem{RS2} R. Bl\"umel, T. M. Antonsen, Jr., B. Georgeot, E. Ott, and
              R. E. Prange,
              Phys. Rev. Lett. {\bf 76}, 2476 (1996);
              Phys. Rev. E {\bf 53}, 3284 (1996).

\bibitem{RS3} L. Sirko, P. M. Koch, and R. Bl\"umel,
               Phys. Rev. Lett. {\bf 78}, 2940 (1997).

\bibitem{RS4} R. Bl\"umel, P. M. Koch, and L. Sirko,
              Found. Phys. {\bf 31}, 269 (2001).

\bibitem{HBohr} H. Bohr, {\it Almost Periodic Functions}
(Chelsea Publishing, New York, 1951).


\bibitem{b10}  R. Artuso, E. Aurell, and P. Cvitanovi\'{c},
               Nonlinearity {\bf 3}, 325 (1990); 361 (1990).

\bibitem{combi} H. Schanz and U. Smilansky, LANL Archive math-ph/0003037.

\bibitem{Sel} A. Selberg, J. Indian Math.\ Soc.\ B {\bf 20},
     47 (1956); reprinted in {\it Atle Selberg: Collected
     Works}, Vol.\ 1 (Springer-Verlag, Berlin, 1989), pp.~423-463.

\end{thebibliography}
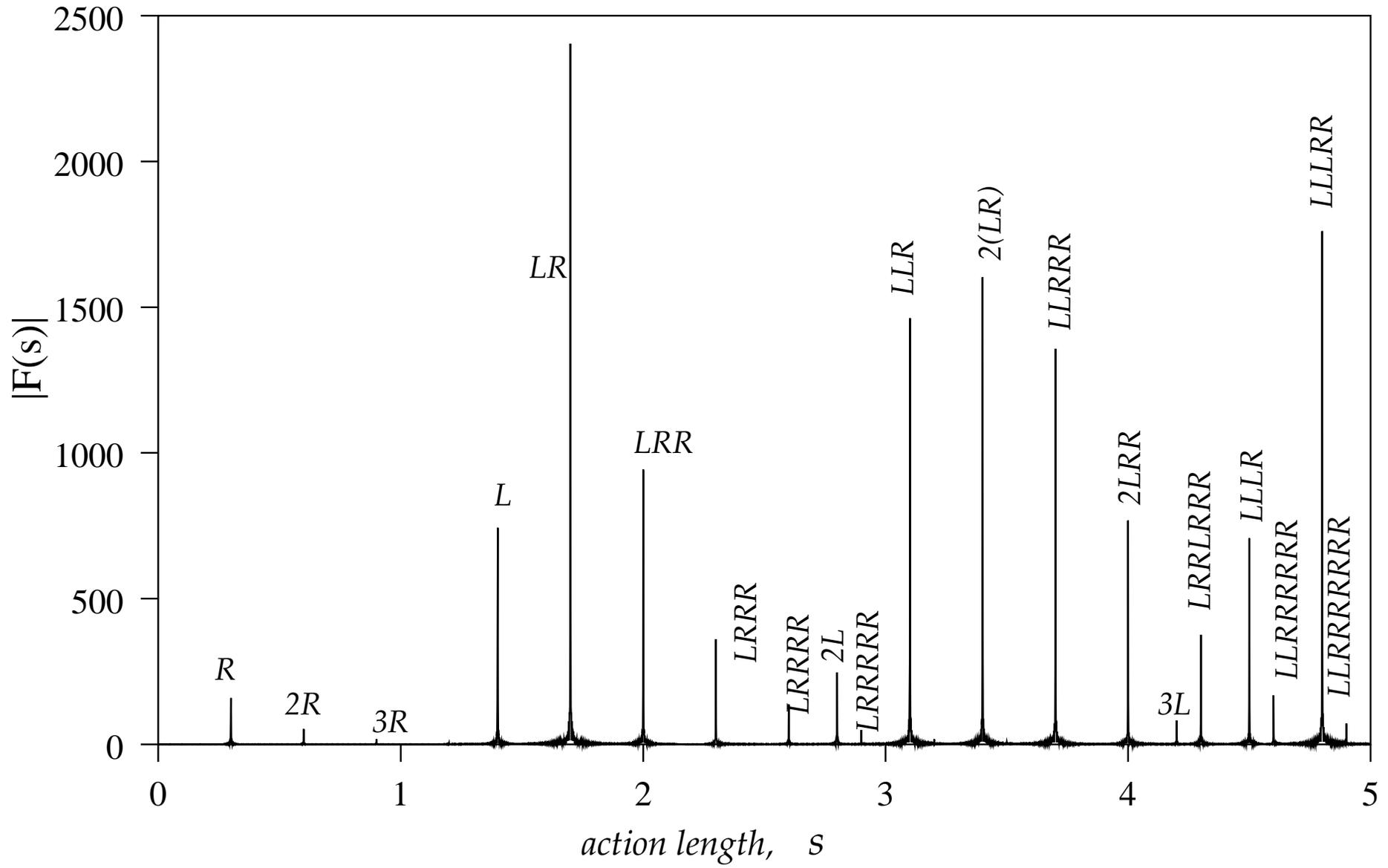
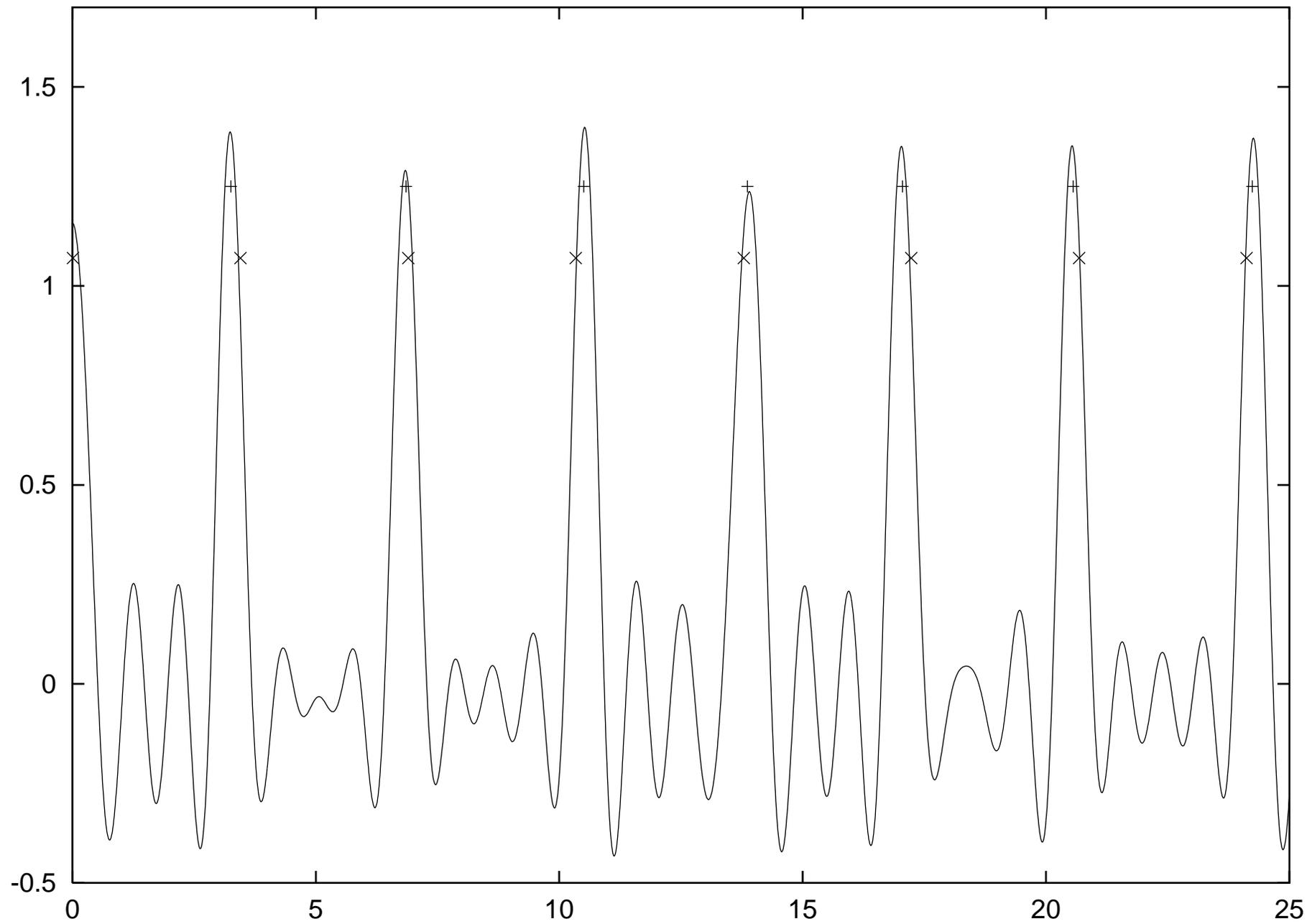
\end{document}